# Topological photonic crystal nanocavity laser

Yasutomo Ota[1*], Ryota Katsumi[2], Katsuyuki Watanabe[1], Satoshi Iwamoto[1,2] and Yasuhiko Arakawa[1]

*Corresponding author
E-mail: ota@iis.u-tokyo.ac.jp

1) Institute for Nano Quantum Information Electronics, The University of Tokyo,
4-6-1 Komaba, Meguro-ku, Tokyo 153-8505, Japan
2) Institute of Industrial Science, The University of Tokyo, 4-6-1 Komaba, Meguro-ku, Tokyo 153-8904, Japan

**Topological edge states exist at the interfaces between two topologically-distinct materials. The presence and number of such modes are deterministically predicted from the bulk-band topologies, known as the bulk-edge correspondence[1]. This principle is highly useful for predictably controlling optical modes[2–5] in resonators made of photonic crystals (PhCs), leading to the recent demonstrations of micro-scale topological lasers[6–10]. Meanwhile, zero-dimensional topological trapped states in the nanoscale remained unexplored, despite its importance for enhancing light-matter interactions and for wide applications including single-mode nanolasers. Here, we report a topological PhC nanocavity with a near-diffraction-limited mode volume and its application to single-mode lasing. The topological origin of the nanocavity, formed at the interface between two topologically-distinct PhCs, guarantees the existence of only one mode within its photonic bandgap[11]. The observed lasing accompanies a high spontaneous emission coupling factor stemming from the nanoscale confinement[12–15]. These results encompass a way to greatly downscale topological photonics[2–5].**

The localization of waves in low dimensions constitutes a basis for diverse applications in various physical systems, including electron, sound and light. Given bulk band topologies, the bulk-edge correspondence, a physical principle that is originally developed for condensed matter physics[1], provides a deterministic route to localizing the waves: the difference in topological invariants between the two materials in contact is associated with the number of existing localized interface modes. Accordingly, zero-dimensional (0D) topological interface states[16], which function as cavities for the waves, can be defined a priori with knowing the bulk band topologies[6,9,10,17–23].

Regarding photonics, in which the control of the number of confined modes is vital for its applications, such deterministic design based on topology[2–5] is highly desired. Indeed, the mode-number control in a micro/nanoscale cavity lacks a rigid strategy and still often relies on empirical approaches. This fact makes a stark contrast with the impressive progress in designing ultra-high $Q$ factor and/or ultra-small mode volume ($V$) nanocavities based on PhCs[24–26] and plasmonics[27]. Recently, topological microcavity lasers designed based on the bulk-edge correspondence have been demonstrated[6–10]. Whilst the topological microcavities exhibited single mode lasing, the cavity designs do not totally deny the existence of other confined modes resonating near the lasing frequencies. Moreover, the spatial confinements of the microscale cavities are relatively weak, hampering the realization of strong light-matter interactions.

In this Letter, we report a topology-based, deterministic design of a single-mode PhC nanocavity and its application to lasing. We show that a 0D edge state is predictably formed between two PhC nanobeams with distinct Zak phases[28], topological invariants in the 1D systems. The 0D edge state functions as a high $Q$ factor nanocavity with a small $V$ close to the diffraction limit. With semiconductor quantum dot gain, the nanocavity exhibits single-mode lasing with a high spontaneous emission coupling factor ($\beta$) of ~ 0.03, which is orders of magnitude larger than those for conventional semiconductor lasers ($\beta$ ~ $10^{-6}$) and can be regarded as a hallmark consequence of the tight optical confinement[12–15].

Figure 1(a) shows a design schematic of the topological PhC nanobeam cavity investigated in this study. Two airbridge PhC nanobeams[15,26] respectively colored in red and blue are interfaced at the cavity center. They share a common 1D photonic bandgap but are distinct in terms of the band topology characterized by the Zak phase, which is defined as an integral of the Berry connection over the first Brillouin zone[11]. Further details of the design are found in Fig. 1(b). The nanobeam is composed of GaAs-based unit cells (refractive index, $n$ = 3.4) patterned with a period of $a$ and is formed with a width of 1.6$a$ and a thickness of 0.64$a$, which are so small that the nanobeam supports a single transverse mode. The unit cells respectively contain two square-shape air holes with a width of 0.5$a$, which are separated by half a period. The two air holes differ in their lengths, $d_1$ and $d_2$, which relate each other via an equation $d_1 + d_2 = 0.5a$. The blue unit cell places the $d_1$ airhole at the center, while the red unit cell has the $d_2$ airhole at the center instead. In the following discussion, we consider situations with $d_1 \leq d_2$, such that $d_1 \leq 0.25a$. The two air holes are arranged within the unit cells so as to keep inversion symmetry, leading to the quantization of Zak phase to either 0 or $\pi$[28]. This allows for predicting the existence of an interface state solely based on the Zak phase[11], which is associated with the symmetry of the relevant bulk band wavefunction[29].

Figure 1(c) show calculated 1D band structures using the 3D plane wave expansion method for the blue and red unit cell when $d_1 = 0.19a$ (left and right panel, respectively) and $d_1 = 0.25a$ (center). At $d_1 = 0.25a$, the two airholes become identical and the nanobeam forms a gapless band structure with a Dirac point at the band edge. Meanwhile, for $d_1 = 0.19a$, a photonic bandgap appears around a normalized frequency of $0.25a/\lambda$, where $\lambda$ is the wavelength of light. The blue and red nanobeam share the same band structure, but the associated wave functions at the band edge have different parities: the blue (red) nanobeam supports a $p(s)$-wave-like magnetic mode with a function node at the unit cell center for the lowest band, while a $s(p)$-wave-like magnetic mode for the second lowest band. Since the optical mode at the zero frequency is always $s$-wave-like, the lowest band for the blue unit cell should have a "twist" in its band nature somewhere in the momentum space. This makes the band topologically-nontrivial in the sense of its finite Zak phase of $\pi$[29], whereas the lowest band for the red unit cell can be regarded as topologically trivial.

Now, we consider a situation where the two 1D PhC nanobeams are put in contact as schematically shown in Figs. 1(a) and (b). When considering the lowest-energy band gap of the 1D system, there predictably exists a single 0D edge state at the interface, when and only when the interface is composed of two 1D PhCs with distinct Zak phases for the lowest bands[11]. Suppose that only $d_1$ and $d_2$ vary in the system, the single-modeness is robustly preserved as long as the two PhCs are topologically distinct, i.e. $d_1 \neq d_2$. The abrupt interface formed with the inversion-symmetric points eliminates the existence of other localized modes (Tamm modes) other than that of the topological origin (Shockley mode[29]). In contrast, the conventional defect PhC nanocavities relying on index modulation sensitively vary the number of confined modes depending on the modulation pattern and in general cannot guarantee the single-modeness[15,26]. Further discussions on the existence of topological edge modes at a variety of interfaces composed of various 1D PhC nanobeams can be found in the supplementary material.

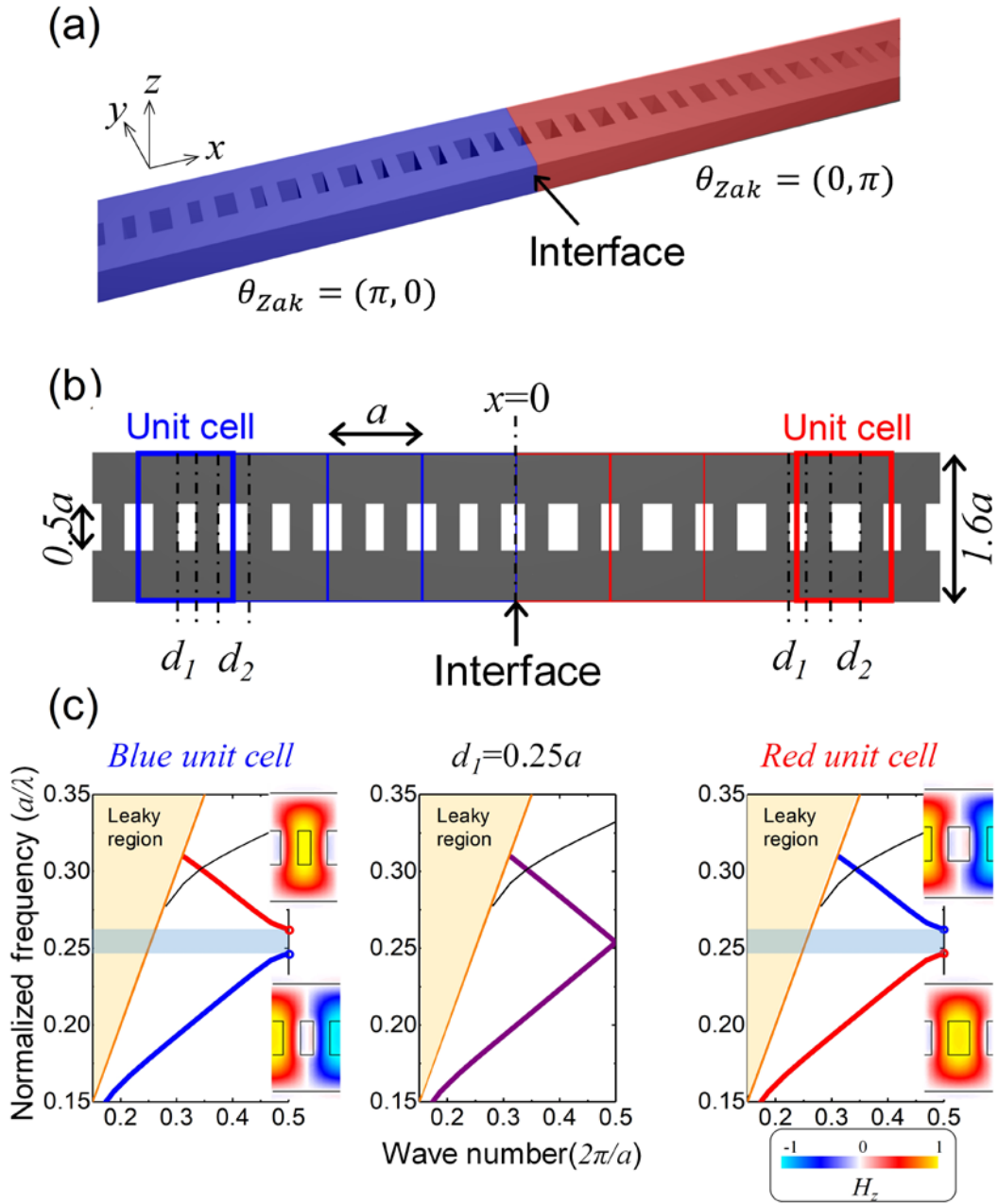

Figure 1. Topological nanocavity design concept. (a) Schematic of the nanocavity formed by interfacing two PhCs with a common bandgap and distinct Zak phases. The Zak phases denoted in the bracket are respectively for the lowest (left) and the second lowest (right) band. (b) Detailed description of the design. The two PhCs are in essence the same but differ in the way to define the unit cells. Blue unit cell arranges the small air hole with a width of $d_1$ at the center, while red unit cell has the large air hole with $d_2$ at the center. The centers of the two holes respectively correspond to the inversion centers of the double periodic PhC. The sum of $d_1$ and $d_2$ equals to $0.5a$. (c) Band structures and wavefunctions at the band edges calculated for the PhC with the blue (left) and red (right) unit cell. The color of the band curve expresses its Zak phase (blue = π, red =0). A band inversion occurs when swapping the unit cells. The center panel shows the case for $d_1 = d_2 = 0.25a$, exhibiting a Dirac point at the band edge.

We simulate the predicted edge modes of the designed nanobeams for different $d_1$ values by the 3D finite difference time domain method. Figure 2(a) shows the calculated electric field profiles of modes appeared within the bandgaps. All the field distributions peak at the interfaces, with the field amplitude decreasing towards the exterior, demonstrating the existence of the localized cavity modes at the interfaces. We numerically confirmed the existence of a single cavity mode within the bandgap of each structure through its excitation spectrum (see the supplementary material).

The spatial extent of the cavity mode depends largely on $d_1$ value, which essentially determines the depth and size of the corresponding bandgap and hence the optical penetration length into the PhC. In our design, the PhC with $d_1 = 0a$ possesses the largest bandgap and thus realizes the tightest optical confinement, resulting in a small mode volume of $0.23(\lambda/n)^3$, which is close to the conventional diffraction limit of $\sim (\lambda/2n)^3$. Figure 2(b) summarizes the evolution of $V$s when increasing $d_1$, showing an exponential-like increase. Meanwhile, interestingly, the calculated resonant frequencies, shown as the red balls in the same plot, do not exhibit a significant dependence on $d_1$[30].

In addition, we calculated the $Q$ factors of the edge modes for different $d_1$ values and plotted them in Fig. 2(c). The $Q$ factor peaks at $d_1 = 0.19a$ and decreases for both higher and lower $d_1$s. For $d_1 \sim 0a$, the optical confinement is too strong to suppress leaky components violating the condition for the total internal reflection, resulting in low $Q$ factors of only about 1,000. When increasing $d_1$ from $0a$, the $Q$ factor improves at the expense of enlarged mode volume. For $d_1 > 0.2a$, $Q$ factor rapidly degrades since the mode size becomes so large compared to that of the allocated simulation domain (70 periods in the $x$ direction), which leads to the light leakage through the domain boundaries. Interestingly, the existence of the 0D edge mode is robust: it resonates even when the patterned PhC area is harshly reduced to containing only a few unit cells[30]. In the current design and simulation size, at $d_1 = 0.19a$, the topological nanocavity support the maximum $Q$ factor of 59,700 with a mode volume of $0.67(\lambda/n)^3$, which are comparable to those for existing PhC nanobeam cavities employed for laser application[15]. The topology-based design enables the wide control of $Q$ factor and mode volume while strictly keeping the single-modeness of the nanocavity. This property is highly beneficial to develop single mode lasers using broadband gain media, including semiconductor quantum dots, since such inhomogeneously-broadened gain media often hamper the selection of a single lasing mode by controlling gain properties.

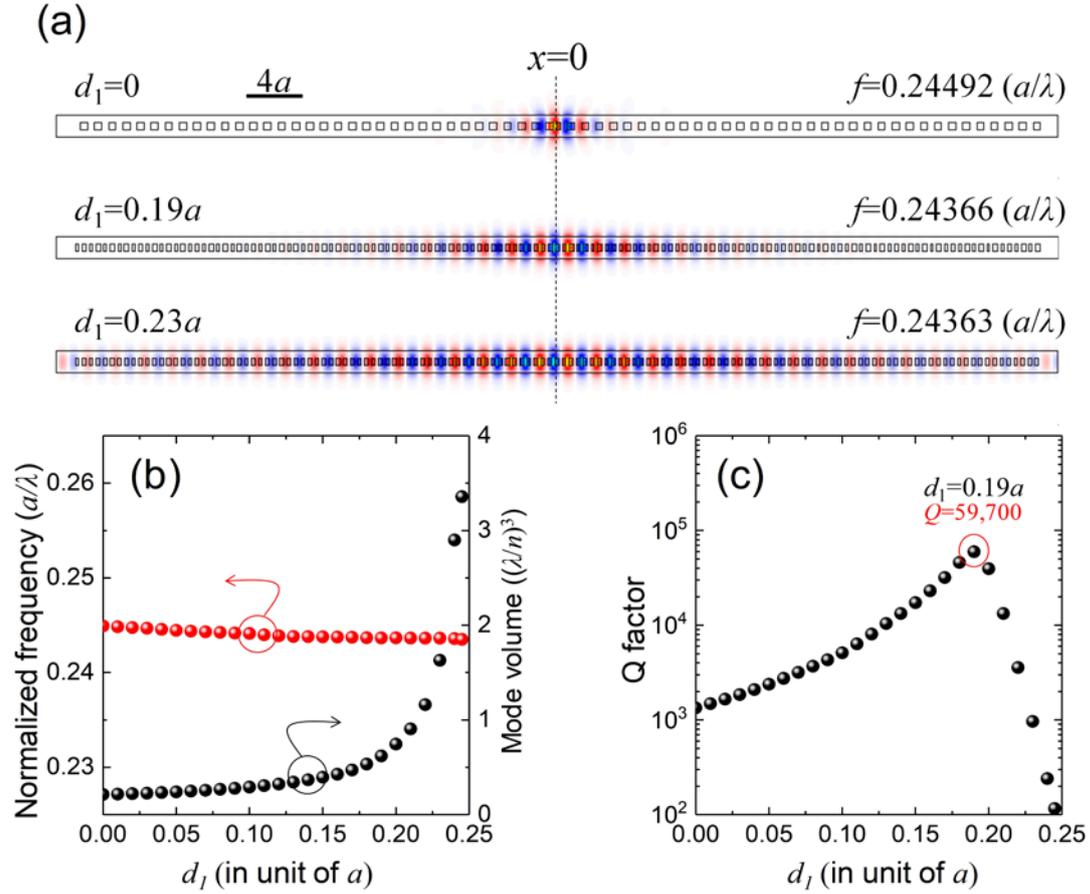

Figure 2. Numerical simulation results. (a) Calculated electric ($E_y$) field profiles for the nanocavities with different $d_1$s. A large modification in the mode extents as $d_1$ increases can be clearly seen. $f$ denotes the normalized frequency of the confined mode. (b) Simulated normalized frequencies (red) and mode volumes (black) plotted as a function of $d_1$. The resonant frequencies are bound closely to the Dirac point, while the mode volumes vary largely as the PhC bandgap narrows by increasing $d_1$. (c) Computed $Q$ factors, showing a peak value of ~ 60,000 at $d_1=0.19a$. The quick decay of $Q$ factor for $d_1 > 0.19a$ originates from the photon leakage from the finite-size simulation region, while the reduction for $d_1 < 0.19a$ is due to the light leakage into free space due to the tighter optical confinement.

We fabricated the designed topological nanocavities with $a$ = 270 nm into a 180-nm-thick GaAs slab containing InAs quantum dots using standard semiconductor nanofabrication processes. The total length of the nanobeam was set to 20 μm (including 68 unit cells). A scanning electron microscope image of a nanocavity designed with $d_1$ = 0.19$a$ is shown in Fig. 3(a). Unlike the theoretical model, the airholes are not perfect squares and possess rounded edges. In the topology-based design, however, the hole shape does not critically influence the existence of the cavity mode as far as the topological bandgap remains opened.

For characterizing the fabricated nanocavity, we performed micro-photoluminescence (µPL) measurements at a low temperature of 10 K. We used a modulated diode laser oscillating at 808 nm (repetition rate 5 MHz, pulse width 20 ns) and an objective lens for optically pumping the samples. PL signals were analyzed with a grating spectrometer and an InGaAs camera. Figure 3(b) shows a series of PL spectra measured with an average pump power of 2.3 µW for the nanocavities designed using different $d_1$s of $0a$, $0.19a$ and $0.23a$. Each spectrum exhibits a single strong peak resonating around a common wavelength of ~1040 nm, as indicated by a colored arrow in the figure. The peak arises from the designed edge mode localized at the interface, confirmed through the experimental evidences provided below. In the figure, we also plotted PL spectra measured for homogeneous PhC nanobeams without the topological interfaces using gray solid lines. The spectra show the frequencies of the band edges of the respective designs. As expected from the design principle, the band gaps that span between the pairs of the band edge peaks actually enclose the emission peaks originated from the interfaces. The measured peak frequencies of the emission peaks are summarized in Fig. 3(c). All the peak positions agree well with those simulated using the plane wave expansion method (solid lines). A highlight of the plot is the nearly-fixed resonant frequencies of the in-gap mode, which are in line with the predicted behavior of the topological edge states. Then, we investigate the spatial extents of the edge modes by measuring the pump position dependences of the PL spectra. The pump spot size in this experiment was about 2 µm. Figure 3(c) shows the normalized peak intensities of the interface modes taken along the $x$ direction. The three curves are of the nanocavities with different $d_1$s and clearly exhibit the mode sizes being dependent on $d_1$. The case of $d_1 = 0a$ and $d_1 = 0.23a$ respectively show the smallest and largest mode distribution, as expected from the simulations. We also characterized the $Q$ factors of the interface modes of different $d_1$s and the corresponding PL spectra are depicted in the right panels in Fig. 3(a). The $Q$ factors for $d_1 = 0a$ and $d_1 = 0.23a$ are measured to be only 700 and 600, respectively. In contrast, we observed a high $Q$ factor of at least over 2,000 for $d_1 = 0.19a$. The measured $Q$ factor for the nanocavity of $d_1 = 0.19a$ is limited by the spectrometer resolution (~400 µeV) used in this particular measurement as well as by the absorption originated from the embedded quantum dots. Therefore, its precise characterization requires further measurements, which we will discuss in the following.

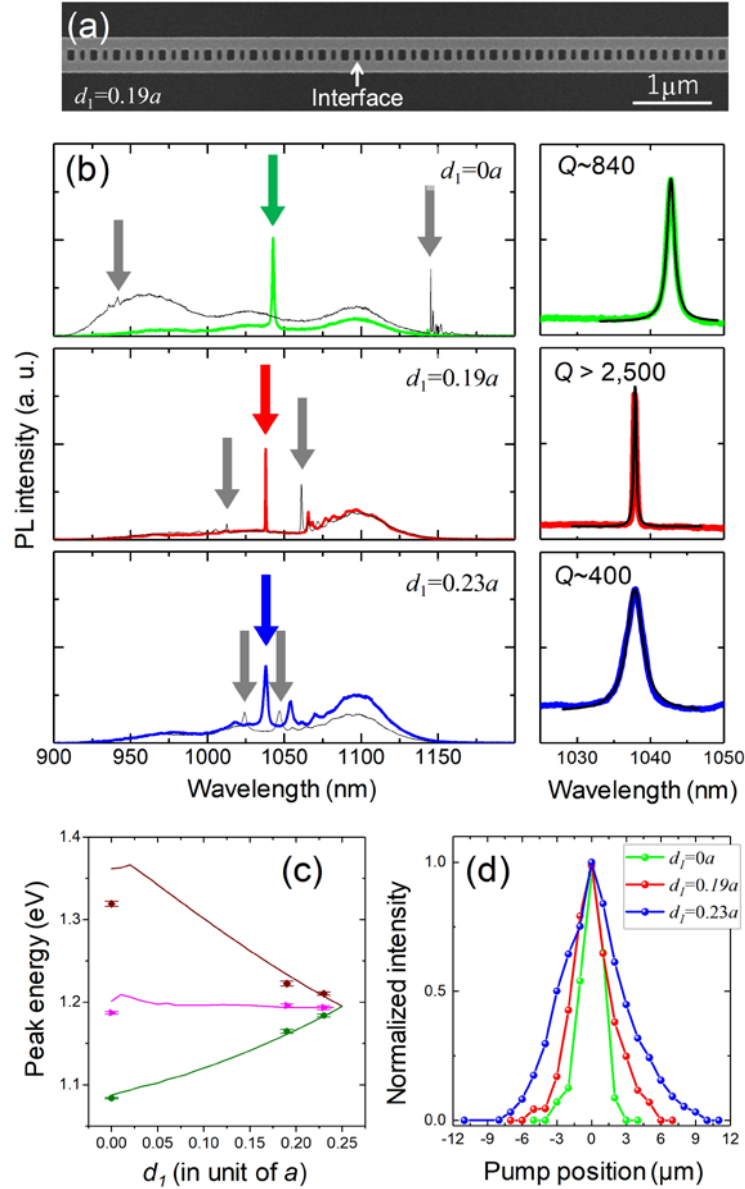

Figure 3. Basic experimental characterization of the topological nanocavities. (a) Scanning electron microscope image of a fabricated nanocavity with $d_1 = 0.19a$. (b) Measured PL spectra for the fabricated PhC nanobeams with different $d_1$s. The colored curves are of topologically-nontrivial interfaces, while the black curves are of regular PhCs. Arrows indicate the spectral peak positions of the topological 0D edge modes (colored) and of the band edges (gray). The right panels show the close up spectra of the topological edge states. The black curves overlaid on the graphs are fitting. (c) Summary of the measured peak positions plotted as a function of $d_1$. The magenta data points are of the 0D confined modes, while the brown (dark green) data points are of the higher (lower) energy band edges. The error bar for each point expresses the standard deviation deduced from the measurement on ten samples with the same design. The solid lines show simulated peak positions by the three dimensional plane wave expansion method. The simulated curves are plotted with a constant offset for better comparison with the experimental results. (d) Position dependent PL intensities measured for the topological nanocavities with different $d_1$s. Tighter mode localization for smaller $d_1$, as expected from the numerical modeling, is demonstrated.

Now, we study the pump power dependence of the topological edge state formed in the $d_1 = 0.19a$ nanocavity, in order to verify its lasing oscillation using the quantum dot gain. For the measurements, we reduced the repetition rate of the pump pulses to 0.5 MHz. Figure 4(a) displays two spectra taken with respective peak pump powers of 5 μW (lower panel) and 150 μW (upper). For the low pump power PL spectrum, we observe broad background emission stemming from the quantum dot spontaneous emission, together with a sharp peak of the localized cavity mode. It is apparent that, with increasing the pump power, the cavity mode emission grows dramatically and dominates the spectrum. In Fig. 4(b), we summarize the peak integrated intensities of the cavity mode measured under various pump powers. The light-in-light-out curve indeed exhibits an intensity jump giving a *s*-like shape, which is typical for lasers and is well fitted using a semiconductor laser model[13]. Through the fit, we deduce a peak threshold pump power of 46 μW and a spontaneous emission coupling factor ($\beta$) of 0.03 for this laser: this high $\beta$ value can be read as a hallmark of nanolasers with tight optical confinement[12–15]. Concomitant to the sharp intensity increase, we observed a clamp of the background emission (see green points in Fig. 4(b)), as another expected phenomenon for lasing. Figure 4(c) shows the evolution of the measured cavity linewidths as a function of pump power, evaluated by using a higher-resolution spectrometer. A significant linewidth narrowing by nearly an order of magnitude is observed, further confirming the lasing oscillation in the investigated topological nanocavity. At a transparency pump power of 11 μw (estimated from the laser model used for the fitting), we estimated a cold cavity $Q$ factor of ~9,600, experimentally demonstrating the formation of a high $Q$ factor mode by the topology-based design. It is noteworthy that the observed lasing can be regarded as quasi-continuous wave since the all the dynamics within the laser is much faster than the pump pulse duration.

In summary, we realized a topological PhC nanocavity laser. We employed the bulk-edge correspondence to deterministically define single mode PhC nanocavities, which support high $Q$ factors and small mode volumes, both of which were confirmed theoretically and experimentally. The tight optical confinement effect manifested itself as the observed high-$\beta$ lasing. Our results showcase an example on how topological physics penetrates into nanophotonics and on how to downscale topological photonics into the nanoscale. We believe that topological photonics[4,5] will renew and advance the understanding and the engineering of nanoscale resonators and lasers.

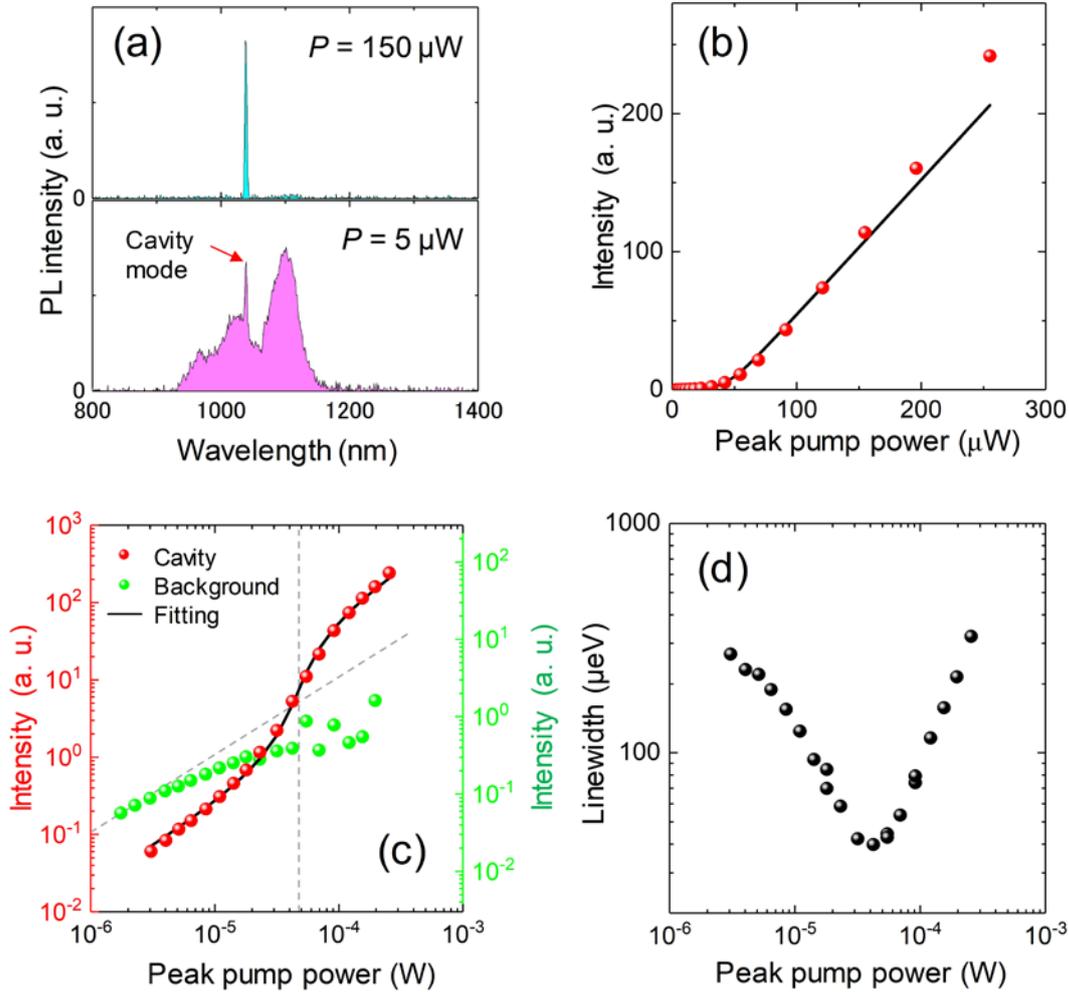

Figure 4. Laser oscillation from the topological nanocavity designed with $d_1 = 0.19a$. (a) PL spectra measured with two different peak pump powers. For the low pumping power of 5 μW, broad background emission from the QDs, spanning from ~ 900 nm to 1200 nm, dominates the spectrum. For the higher pumping power, the background is suppressed and the cavity emission peak in turn prevails. (b) Light-in-light-out plot for the emission peak of the 0D edge state. The black solid line is the fitting to the data points. Laser transition with a threshold peak pump power of 46 μW is seen. (c) Logarithmic plots of the LL curve (red) and of the evolution of the intensities of the background emission (light green). The LL curve exhibits a mild S-shaped curve, which is typical for high-$β$ lasers. The vertical dashed line indicates the threshold pump power. The background emission clamps above the laser threshold. The diagonal dashed line is an eye guide with a linear increase. The scattered data points above the threshold originate from the noisy PL spectra for the QD emission, due to strong cavity emission and the limited dynamic range of our detector. (d) Measured evolution of cavity linewidths as a function of peak pump power. The linewidths were extracted by fitting to the high resolution spectra with Lorentzian peak functions convolved with a Gaussian function representing the spectrometer response (~40 μeV). Linewidths shows a significant narrowing of nearly an order of magnitude. The increase of the linewidth above the lasing threshold is likely due to heating in the nanocavity.

Acknowledgments

The authors thank Y. Hatsugai, C. F. Fong and I. Kim for fruitful discussions. This work was supported by JSPS KAKENHI Grant-in-Aid for Specially promoted Research (15H05700), KAKENHI(17H06138, 15H05868), JST-CREST and was based on results obtained from a project commissioned by the New Energy and Industrial Technology Development Organization (NEDO).

References

1. Hatsugai, Y. Chern number and edge states in the integer quantum Hall effect. *Phys. Rev. Lett.* **71,** 3697–3700 (1993).
2. Haldane, F. D. M. & Raghu, S. Possible Realization of Directional Optical Waveguides in Photonic Crystals with Broken Time-Reversal Symmetry. *Phys. Rev. Lett.* **100,** 013904 (2008).
3. Wang, Z., Chong, Y., Joannopoulos, J. D. & Soljačić, M. Observation of unidirectional backscattering-immune topological electromagnetic states. *Nature* **461,** 772–5 (2009).
4. Lu, L., Joannopoulos, J. D. & Soljačić, M. Topological photonics. *Nat. Photonics* **8,** 821–829 (2014).
5. Ozawa, T. *et al.* Topological Photonics. *arXiv:1802.04173* (2018).
6. St-Jean, P. *et al.* Lasing in topological edge states of a one-dimensional lattice. *Nat. Photonics* **11,** 651–656 (2017).
7. Bahari, B. *et al.* Nonreciprocal lasing in topological cavities of arbitrary geometries. *Science (80-. ).* **358,** 636–640 (2017).
8. Bandres, M. A. *et al.* Topological insulator laser: Experiments. *Science (80-. ).* **359,** eaar4005 (2018).
9. Parto, M. *et al.* Edge-Mode Lasing in 1D Topological Active Arrays. *Phys. Rev. Lett.* **120,** 113901 (2018).
10. Zhao, H. *et al.* Topological hybrid silicon microlasers. *Nat. Commun.* **9,** 981 (2018).
11. Xiao, M., Zhang, Z. Q. & Chan, C. T. Surface Impedance and Bulk Band Geometric Phases in One-Dimensional Systems. *Phys. Rev. X* **4,** 021017 (2014).
12. Khajavikhan, M. *et al.* Thresholdless nanoscale coaxial lasers. *Nature* **482,** 204–207 (2012).
13. Ota, Y., Kakuda, M., Watanabe, K., Iwamoto, S. & Arakawa, Y. Thresholdless quantum dot nanolaser. *Opt. Express* **25,** 19981 (2017).

14. Takiguchi, M. *et al.* Systematic study of thresholdless oscillation in high-β buried multiple-quantum-well photonic crystal nanocavity lasers. *Opt. Express* **24,** 3441 (2016).
15. Zhang, Y. *et al.* Photonic crystal nanobeam lasers. *Appl. Phys. Lett.* **97,** 051104 (2010).
16. Su, W. P., Schrieffer, J. R. & Heeger, A. J. Solitons in polyacetylene. *Phys. Rev. Lett.* **42,** 1698–1701 (1979).
17. Gao, W. S., Xiao, M., Chan, C. T. & Tam, W. Y. Determination of Zak phase by reflection phase in 1D photonic crystals. *Opt. Lett.* **40,** 5259–62 (2015).
18. Tan, W., Sun, Y., Chen, H. & Shen, S.-Q. Photonic simulation of topological excitations in metamaterials. *Sci. Rep.* **4,** 3842 (2015).
19. Poli, C., Bellec, M., Kuhl, U., Mortessagne, F. & Schomerus, H. Selective enhancement of topologically induced interface states in a dielectric resonator chain. *Nat. Commun.* **6,** 6710 (2015).
20. Slobozhanyuk, A. P., Poddubny, A. N., Miroshnichenko, A. E., Belov, P. A. & Kivshar, Y. S. Subwavelength Topological Edge States in Optically Resonant Dielectric Structures. *Phys. Rev. Lett.* **114,** 123901 (2015).
21. Xiao, M. *et al.* Geometric phase and band inversion in periodic acoustic systems. *Nat. Phys.* **11,** 240–244 (2015).
22. Xiao, Y.-X., Ma, G., Zhang, Z.-Q. & Chan, C. T. Topological Subspace-Induced Bound State in the Continuum. *Phys. Rev. Lett.* **118,** 166803 (2017).
23. Kim, I., Iwamoto, S. & Arakawa, Y. Topologically protected elastic waves in one-dimensional phononic crystals of continuous media. *Appl. Phys. Express* **11,** 017201 (2018).
24. Asano, T., Ochi, Y., Takahashi, Y., Kishimoto, K. & Noda, S. Photonic crystal nanocavity with a Q factor exceeding eleven million. *Opt. Express* **25,** 1769 (2017).
25. Hu, S. *et al.* Experimental Realization of Deep Subwavelength Confinement in Dielectric Optical Resonators. *arXiv:1707.04672* (2017).
26. Deotare, P. B., McCutcheon, M. W., Frank, I. W., Khan, M. & Lončar, M. High quality factor photonic crystal nanobeam cavities. *Appl. Phys. Lett.* **94,** 121106 (2009).
27. Oulton, R. F. *et al.* Plasmon lasers at deep subwavelength scale. *Nature* **461,** 629–32 (2009).
28. Zak, J. Berrys phase for energy bands in solids. *Phys. Rev. Lett.* **62,** 2747–2750 (1989).
29. Zak, J. Symmetry criterion for surface states in solids. *Phys. Rev. B* **32,** 2218–2226 (1985).

30. Kalozoumis, P. A. *et al.* Finite size effects on topological interface states in one-dimensional scattering systems. *arXiv:1712.08763* (2017).

# Supplementary material : Topological photonic crystal nanocavity laser

Yasutomo Ota[1*], Ryota Katsumi[2], Katsuyuki Watanabe[1], Satoshi Iwamoto[1,2] and Yasuhiko Arakawa[1]

*Corresponding author
E-mail: ota@iis.u-tokyo.ac.jp

1) Institute for Nano Quantum Information Electronics, The University of Tokyo,
4-6-1 Komaba, Meguro-ku, Tokyo 153-8505, Japan
2) Institute of Industrial Science, The University of Tokyo, 4-6-1 Komaba, Meguro-ku, Tokyo 153-8904, Japan

**1. The existence or absence of confined edge modes at the interfaces defined by several combinations of 1D PhC nanobeams.**

We analyzed several combinations of 1D PhC nanobeams using the 2D FDTD method. We consider two types of unit cells for the 1D PhC nanobeams, as shown in Fig. S1. The A-type is in essence the same as that in the main text and is composed of two air holes with different sizes. The width of the center air hole is $d_1$, while that of the other is $0.25a - 0.5d_1$. The B-type contains two airholes with a width of $0.25a$ that are separated by $L$. This unit cell can be regarded as a complementary structure of the A-type in the sense that one becomes the other after the conversion of airholes to dielectric and vice versa when $L - d_1 = 0.25a$. The average refractive indexes of the two unit cells are the same, as both the structures contain air region with a total area of $0.5a \times 0.8a$. Both the unit cells are symmetric under the inversion operation at the center, such that the Zak phases of photonic bands are quantized to be either 0 or π. We choose the lattice constant ($a$) to be 270 nm and the refractive index of the nanobeam dielectric material to be 2.7.

We investigated the existence and absence of interface modes for various combinations of 1D PhC nanobeams composed of A and/or B type unit cells, as summarized in Fig. S2. In the simulations, we excite a non-symmetric point near the interface of each 1D PhC nanobeam with a current impulse and record the time evolutions of Ex, Ey and Hz fields at various locations in the simulator. By Fourier transforming the recorded time evolutions, we obtained excitation spectra of the 1D PhC nanobeam. Typically, we observed a bandgap centered at around 0.84 μm$^{-1}$. When we find an in-gap peak in the excitation spectrum, we analyzed its origin by selectively exciting it and computing its mode distribution in the simulator.

The situation (1) in Fig. S2 corresponds to the one discussed in the main text. The unite cells originate from the same 1D PhC but are defined using different inversion-symmetric points that are separated by $a/2$. In this case, photonic bands in the PhC based on the two unit cells are inverted and the associated Zak phases always differ by π. Based on the bulk-edge correspondence discussed in the main text, the interface deterministically supports a single 0D edge state. This can be confirmed in the excitation spectrum: a single prominent peak is observed within the bandgap. The localization of the in-gap mode at the vicinity of the interface is confirmed by the simulation of the mode distribution.

The situation (2) differs from (1) in the hole sizes of the right-side 1D PhC. Now the existence of the interface itself becomes clearer. However, the two PhCs are the same in the sense of topology, i.e. sharing the same Zak phases for the lowest energy bands. In this case, the bulk edge correspondence predictably denies the existence of an edge mode. Indeed, the excitation spectra exhibit a feature-less bandgap region around a frequency of 0.84 μm$^{-1}$.

The situation (3) differs from (2) in the hole sizes of the left-side PhC nanobeam. We choose a bigger $d_1$ for the left-side PhC such that the bandgap size shrinks. Indeed, the corresponding excitation spectra exhibit a narrowed bandgap compared to those in (1) and (2). Nevertheless, a single interface mode exists within the gap thanks to the different Zak phases between the two constituent PhCs. Due to the reduced depth of the bandgap in the left-side PhC and hence the reflectivity by the single unit cell, the localized mode penetrates more largely into the left-side PhC.

The situation (4) differs from (1) and (2) in the type of unit cell of the right-side PhC nanobeam. Being a complementary structure of the A-type unit cell, the newly-introduced B-type unit cell on the right-side PhC fulfills $L - d_1 = 0.25a$. In this case, the complementary B-type cell supports an inverted lowest-energy band structure than that of the A-type, leading to the difference in the Zak phase. Accordingly, we observe a single interface 0D localized mode within the bandgap.

The situation (5) differs from (4) in the length of $L$ of the right-side PhC. By increasing $L$ over the topological transition point of $L = 0.5a$, the band inversion occurs. Now the Zak phases for the lowest energy bands of the left- and right-side PhCs are the same, leading to the absence of interface modes. Indeed, we did not observe any clear peak within the bandgap.

The situation (6) is a complementary version of (1), generated from replacing dielectric and airholes for both the right- and left-side 1D PhCs in the situation of (1). The topological natures for both the PhCs also invert, resulting in the deterministic 0D edge state at the interface.

Overall, we observed the deterministic formation of a single 0D edge state at the interfaces constituted by two topologically-distinct PhC nanobeams. The interface states formed within the lowest-energy bandgaps emerge only when the Zak phases for the lowest-energy bands of the two PhCs differ. The topology-based design of the nanocavities provides a novel pathway for controlling the number of optical modes.

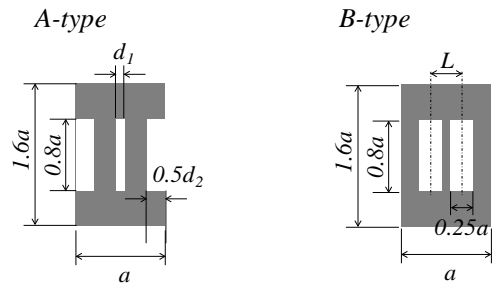

Figure S1. Unit cells under consideration

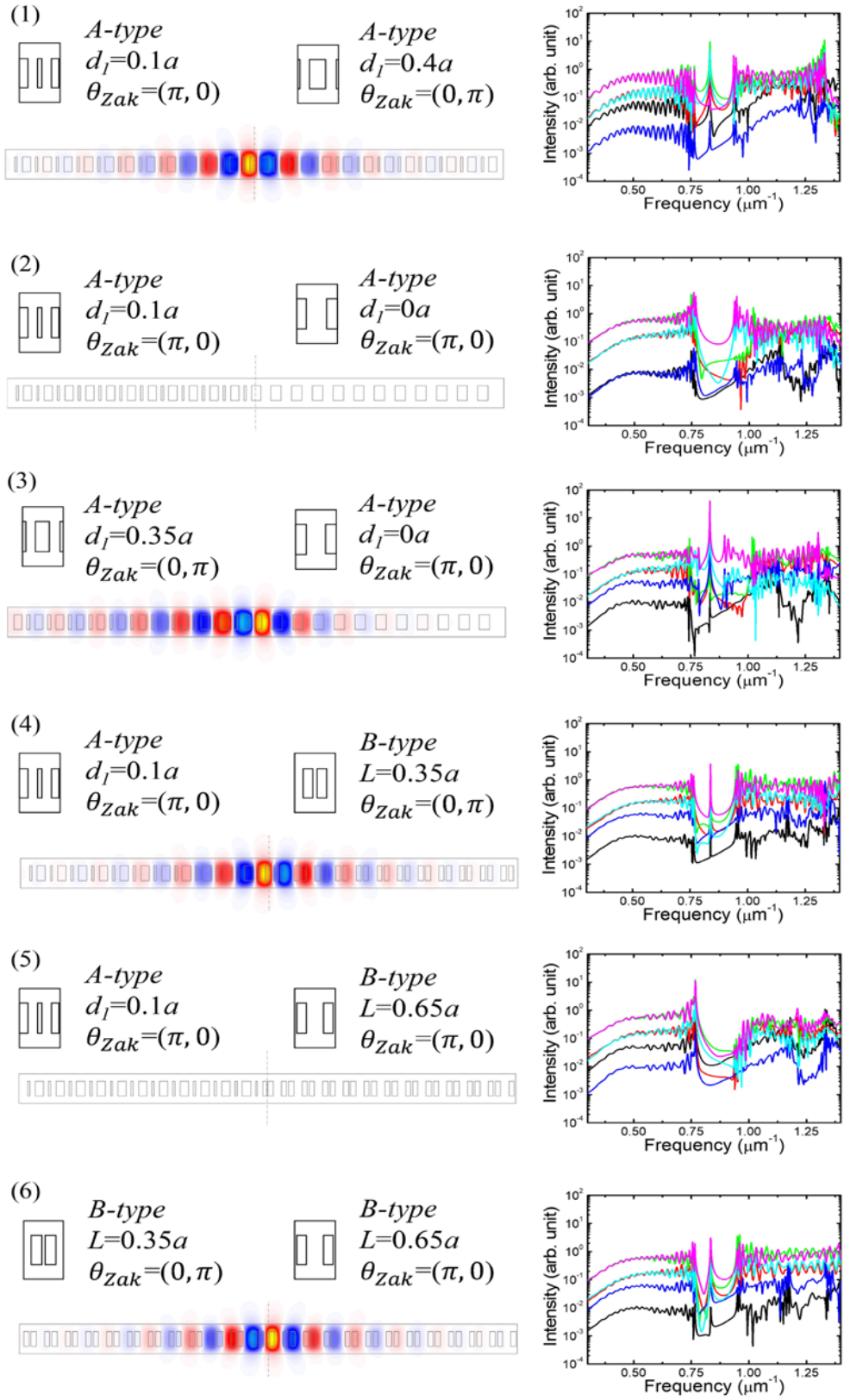

Figure S2. Existence and absence of topological interface states at the interfaces between several different combinations of 1D PhC nanobeams